\documentclass[11pt,english,a4paper]{article}
\usepackage{graphicx}
\usepackage{authblk}
\usepackage{wrapfig}
\usepackage{amsmath}
\usepackage{geometry} 
\usepackage{typearea}
\typearea{10}

\newcommand{\dru}{day$^{-1}$kg$^{-1}$keV$^{-1}$ }
\newcommand{\kevee}{keV$_\mathrm{ee}$ }
\newcommand{\znu}{$0\nu\beta\beta$}
\newcommand{\tnu}{$2\nu\beta\beta$}

\title{
{\Large
PRESENT STATUS OF PICOLON PROJECT}
}

\author{K.~Fushimi$^{*}$}
\affil{Division of Science and Technology, Tokushima University, 2-1 Minami Josanajima-cho, Tokushima city, Tokushima , 770-8506, Japan}
\author{D.~Chernyak}
\affil{Department of Physics and Astronomy, University of Alabama, Tuscaloosa, Alabama 35487, USA}
\author{H.~Ejiri}
\affil{Research Center for Nuclear Physics, Osaka University, 10-1 Mihogaoka Ibaraki city, Osaka, 567-0042, Japan}
\author{K.~Hata}
\affil{Research Center for Neutrino Science, Tohoku University, 6-3 Aramaki Aza Aoba, Aobaku, Sendai city, 
Miyagi, 980-8578, Japan }
\author{R.~Hazama}
\affil{Department of Environmental Science and Technology, Osaka Sangyo University, 3-1-1 Nakagaito, 
Daito city, Osaka, 574-8530, Japan}
\author{T.~Iida}
\affil{Faculty of Pure and Applied Sciences, University of Tsukuba, 1-1-1 Tennoudai, Tsukuba city, 
Ibaraki, 305-8571, Japan}
\author[5]{H.~Ikeda}
\author{K.~Imagawa}
\affil{I.~S.~C. Lab.~, 5-15-24 Torikai Honmachi, Settsu city, Osaka, 566-0052, Japan}
\author[5,8]{K.~Inoue}
\affil{Kavli Institute for the Physics and Mathematics of the Universe (WPI),
5-1-5 Kashiwanoha, Kashiwa city, Chiba, 277-8583, Japan}
\author{H.~Ito}
\affil{Department of Physics, Faculty of Science and Technology, Tokyo University of Science, Noda, Chiba 278-8510, Japan}
\author{T.~Kisimoto}
\affil{Department of Physics, Osaka University, 1-1 Machikaneyama-cho,
Toyonaka city,  Osaka 560-0043, Japan}
\author[5,8]{M.~Koga}
\author{K.~Kotera}
\affil{Graduate School of Advanced Technology and Science, Tokushima University, 2-1 Minami Josanajima-cho, 
Tokushima city, Tokushima , 770-8506, Japan}
\author{A.~Kozlov}
\affil{National Research Nuclear University ``MEPhI'' (Moscow Engineering Physics Institute), Moscow, 115409, Russia}
\author[13,14,15]{S.~Kurosawa}
\affil{New Industry Creation Hatchery Center, Tohoku University, 6-6-10 Aza-Aoba, Aramaki, Aoba-ku, Sendai, 980-8579, Japan}
\affil[14]{Institute for Materials Research, Tohoku University, 2-1-1 Katahira, Aoba-ku, Sendai, Miyagi 980-8577, Japan}
\affil[15]{Institute of Laser Engineering, Osaka University, 2-6 Yamadaoka, Suita, Osaka 565-0871, Japan}
\author[8,16]{K.~Nakamura}
\affil[16]{Butsuryo College of Osaka, 3-33 Ohtori Kitamachi, Nishi ward, Sakai city, Osaka, 
593-8328, Japan }
\author[1]{R.~Orito}
\author[6]{A.~Sakaguchi}
\author[11]{A.~Sakaue}
\author[4]{T.~Shima}
\author[6]{Y.~Takaku}
\author[8,15]{Y.~Takemoto}
\affil[17]{Institute for Cosmic Ray Research, The University of Tokyo, 5-1-5 Kashiwanoha, Kashiwa city, Chiba, 277-8583, Japan}
\author[4]{S.~Umehara}
\author[11]{Y.~Urano}
\author[1]{Y.~Yamamoto}
\author[7]{K.~Yasuda}
\author[10]{S.~Yoshida}

\date{}
\begin{document}

\maketitle
\normalsize
\begin{abstract}
The existence of cosmic dark matter and neutrino properties are long-standing problems in 
cosmology and particle physics.
These problems have been investigated by using radiation detectors.
We will discuss the application of inorganic crystal scintillators to studies on dark matter and 
neutrino properties.
A large volume and high-purity inorganic crystal is a promising detector for investigating dark matter and neutrino.
\end{abstract}

\section{Introduction}
\subsection{Dark matter problem}

Cosmic dark matter is an unknown elementary particle that does not interact via electromagnetic interaction.
There are two attractive candidates from particle physics.
One is weakly interacting massive particles (WIMPs), and the other is the axion.

\begin{figure}[hb]
\centering
\includegraphics[width=\linewidth]{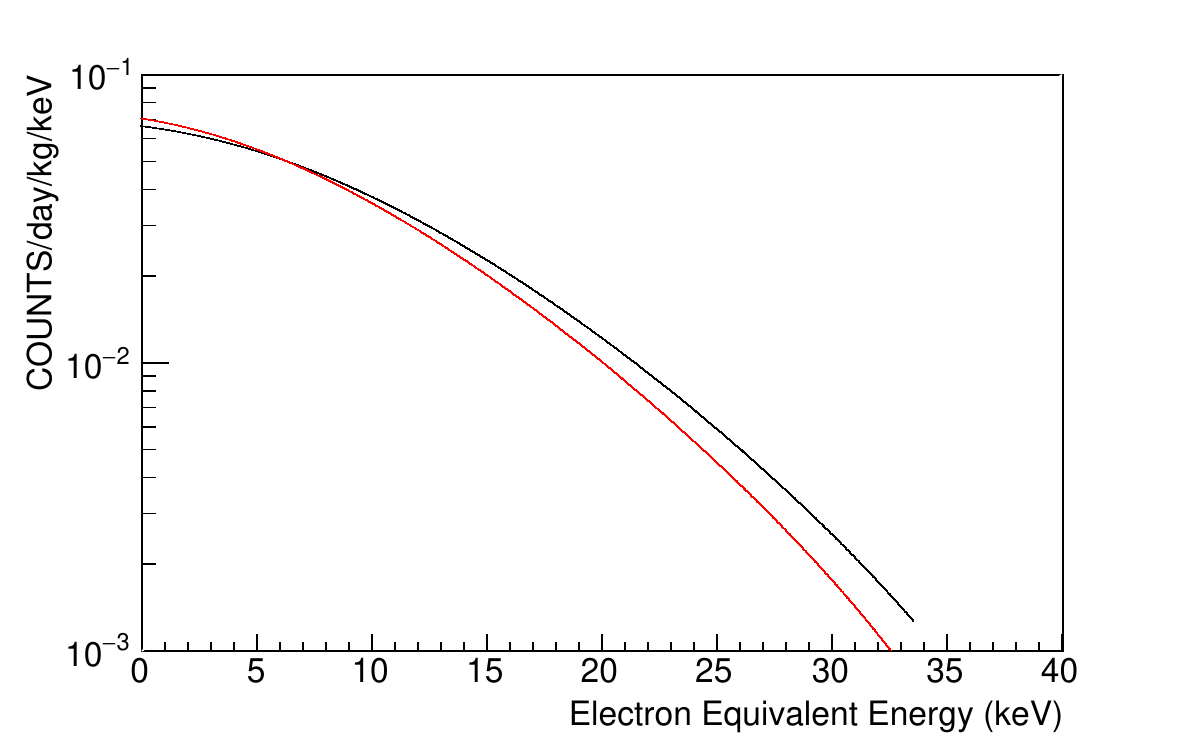}
\caption{The expected energy spectra of recoil Na in NaI(Tl) scintillator.
The black and red lines are the recoil energy spectra observed in June and December.}
\label{fg:spe_Na}
\end{figure}

WIMPs are a sort of hypothetical elementary particles which interact via 
weak interaction.
Their mass ranges from sub-GeV to a few TeV.\@
The root mean square of the WIMPs velocity is 238~km/s \cite{Sofue2020, Honma2012, Honma2015}.
It gives recoil energy to an atomic nucleus by elastic scattering.
The mean recoil energy and event rate are modulated annually because of the proper motion of the sun and the earth's revolution.
The maximum of WIMPs energy is given at the beginning of June,
and the minimum is given at the beginning of December \cite{Freese1988, PhysRevD.33.3495}.
Figure \ref{fg:spe_Na} shows the expected energy spectrum of Na nuclei in a NaI(Tl) scintillator recoiled by WIMPs with its mass of 50~GeV$/c^{2}$, where $c$ is the speed of light.
The energy spectra were calculated with the cross-section of elastic scattering between proton and WIMPs as
$\sigma_{\mathrm{p}-\chi}=10^{-36}$~cm$^{2}$.

Axion is another candidate for dark matter.
It was proposed to solve the CP violation problem in strong interaction \cite{Peccei1977, Weinberg1978}.
The axion is a light boson and interacts via the Peccei-Quinn mechanism, which is analogous to photoelectronic interaction.

Many experimental works have been done to detect dark matter candidates by using various radiation detectors.
The semiconductor detector has a significant advantage in particle identification.
The ratio of phonons to ionization is small for electrons and large for recoil nuclei\cite{PhysRevLett.112.241302}.
The semiconductor detectors can perform almost background-free experiments since the WIMPs signal gives nuclear recoil signal.

The other detector, which gives excellent results, is a large-volume xenon scintillator \cite{CLINE2000373, PhysRevLett.129.161805}.
The two-phase xenon detector not only discriminates electron-recoil and nuclear recoil but also identifies the position of the interaction.
XENONnT detector combines a scintillation detector and a TPC (time projection chamber) detector.
TPC measures the ionization signal and the position of interaction.
The signal intensity ratio between ionization and scintillation enables discriminating between nuclear and electron recoil events.
Recently, XENONnT group reported the lowest background level in electron recoil events as $(4.3\pm0.4)\times 10^{-5}$~\dru by $5.9$~ton liquid xenon detector \cite{PhysRevLett.129.161805}.
Despite the low-background detectors, no significant signal of WIMPs has been reported.
The experimental limit on the cross-section was dug down to $10^{-45}$~cm$^{2}$.
We need to develop a low-background detector as low as
$10^{-5}$~\dru.

\subsection{Neutrino mass}
The neutrino mass is not measured directly, although the evidence of a finite mass was reported by neutrino oscillation.
The neutrinos have three flavor eigenstates in weak interaction, $\nu_{\mathrm{e}}$, $\nu_{\mu}$, and $\nu_{\tau}$.
If the neutrinos have finite mass, the flavor eigenstate is expressed by a linear combination of the mass eigenstate, $\nu_{1}$,
$\nu_{2}$, and $\nu_{3}$.
The relation between flavor eigenstate and mass eigenstate is expressed as
\begin{equation}
\begin{pmatrix}
\nu_{\mathrm{e}} \\
\nu_{\mu} \\
\nu_{\tau} \\
\end{pmatrix} =
\begin{pmatrix}
U_{\mathrm{e}1} & U_{\mathrm{e}2} & U_{\mathrm{e}3} \\
U_{\mu 1} & U_{\mu 2} & U_{\mu 3} \\
U_{\tau 1} & U_{\tau 2} & U_{\tau 3} \\
\end{pmatrix}
\begin{pmatrix}
\nu_{1} \\
\nu_{2} \\
\nu_{3} \\
\end{pmatrix}
\end{equation}

The observable eigenstate of the neutrino is its weak eigenstate because we observe them via weak interaction with electrons and muons.
The neutrino flavor changes traveling between the emission point and
detection point.
The probability of detecting an electron-neutrino at a distance depends on the difference in squared mass
\begin{equation}
\Delta m^{2}_{ij} = m^{2}_{i}-m^{2}_{j},
\end{equation}
where $m_{i}, i=1,2,3$ is the mass of the neutrinos $\nu_{i}, i=1,2,3$.
The detection probability of the original neutrino flavor oscillates with the distance of the neutrino source, called neutrino oscillation.
The neutrino oscillation insists that the neutrino has a finite mass. However, the experimental results do not tell its absolute value.

The neutrino-less double beta decay (\znu ) is one of the most sensitive methods to measure the absolute mass of the neutrino \cite{EjiriJPSJ, Jones2021}.
Double beta decay occurs in the nuclei if single beta decay is prohibited and double beta decay is allowed.
The nucleons that make up the nucleus are arranged in order from the lowest energy orbital. Two nucleons in the same orbit settle in a lower energy state when their spins are opposite.
Nuclei with even numbers of protons and neutrons, called even-even nuclei, have less mass than odd-odd nuclei with an odd number of both. As a result, $^{136}$Cs next to $^{136}$Xe is heavier than $^{136}$Xe, and $^{136}$Ba next to it is lighter than $^{136}$Xe.
This energy difference causes double beta decay.

\begin{figure}[h]
\centering
\includegraphics[width=\linewidth]{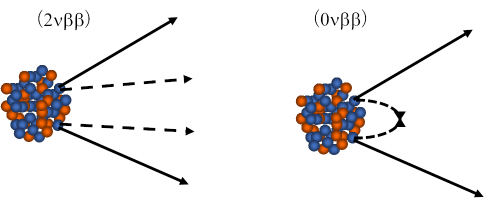}
\caption{Image of the double beta decay. The neutron and proton in a nucleus are blue and red circles.
A solid line and a dashed line stand for the electron and neutrino, respectively}
\label{fg:DBD}
\end{figure}

The two-neutrino double beta decay (\tnu ) is expressed by the decay formula as
\begin{equation}
(A, Z)\rightarrow (A,Z+2)+2\mathrm{e}^{-}+\bar{\nu_{\mathrm{e}}},
\end{equation}
where $A$ and $Z$ are the nucleus's mass number and atomic number (Left of Figure \ref{fg:DBD}).
The \tnu \ occurs in the flame of the standard model of particle physics.

An electron has two chirality states, right-handed ($\mathrm{e}_{\mathrm{R}}$) and left-handed ($\mathrm{e}_{\mathrm{L}}$).
Chirality is defined as
\begin{eqnarray}
\psi_{\mathrm{L}} &= \frac{1-\gamma^{5}}{2}\psi \\
\psi_{\mathrm{R}} &= \frac{1+\gamma^{5}}{2}\psi.
\end{eqnarray}
The four-component Dirac spinor $\psi$ is written in terms of two upper components of the left chirality field $\psi_{\mathrm{L}}$ and two lower components of the right chirality field $\psi_{\mathrm{R}}$.
The charged leptons have two chirality states, L and R.\@
On the other hand, neutrinos have only left-handed chirality.
Therefore, only left-handed neutrino $\nu_{\mathrm{L}}$ and right-handed antineutrino $\bar{\nu_{\mathrm{R}}}$ exist
in the standard model.

\znu \ process gives a significant information to the beyond standard model.
\znu \ process occurs if the neutrino has right-handed chirality and the neutrino and anti-neutrino are the same, called Majorana neutrino.
The \znu \ is expressed as
\begin{equation}
(A, Z)\rightarrow (A,Z+2)+2\mathrm{e}^{-}.
\end{equation}
In this case, a neutrino emitted by one neutron is absorbed by another in the same nucleus or annihilates each other (see right side of Figure \ref{fg:DBD}).
The decay probability of \znu \ is expressed by Majorana mass $\left< m \right>$, and right-handed interaction in terms of $\left<\lambda \right>$ and $\left< \eta \right>$.
The half-life of \znu \ decay, $T^{0\nu}_{1/2}$, is expressed as
\begin{equation}
\frac{\ln 2} {T^{0\nu}_{1/2}}= 
G^{0\nu}\left| M^{0\nu}\right|^{2}K,
\end{equation}
where $G^{0\nu}$ is the kinematical factor and $M^{0\nu}$ is the
nuclear matrix element.
The kinematical factor and nuclear matrix element are obtained by nuclear physics \cite{EjiriJPSJ, Jones2021}.
$K$ is the neutrino mass term $\left<m_{\nu}\right>$, 
and the right-handed current terms 
$\left<\lambda\right>$ and $\left<\eta\right>$.

The expected half-life of \znu \ is more than $10^{6}$ longer than that of \tnu \ process.
KamLAND-Zen group searched for \znu \ of $^{136}$Xe by 970~kg$\cdot$yr exposure \cite{KL-Zen_PRL2023}.
They obtained a lower limit on the half-life of \znu \ decay  of 
$T_{1/2}^{0\nu} > 2.3\times 10^{26}$~yr at 90~\% C.L.\@
Their limit corresponds to upper limits on the effective Majorana 
neutrino mass of $36\sim 156$~meV.\@
The mass range of the effective Majorana neutrino comes from the nuclear matrix elements calculated by several models.
The lower half-life limits by other groups are longer than the order of $10^{25}$~yr.
One needs to develop an extremely low background radiation detector to search for \znu \ events.

\section{PICOLON Project}
PICOLON (Pure Inorganic Crystal Observatory for LOw-energy Neutr(al)ino) is a research project
to search for WIMPs, and double beta decays, especially \znu\, using extremely pure inorganic crystals.
We are currently developing detectors for the WIMP search using NaI(Tl)\cite{FushimiISRD2016, FushimiPTEP2020}.

We have established a technique to remove radioactive impurities in NaI(Tl) crystals.
Table \ref{tb:contami} shows our recent results and comparisons with other groups applying NaI(Tl) crystal.
We calculated the U-series contamination from $^{226}$Ra since this series is divided into five sub-series by long-lived progeny.
The $^{226}$Ra produces many isotopes, which emit many beta-rays and gamma-rays.
The beta- and gamma-rays from them are supposed to be the primary origin of the present background.

\begin{table}[ht]
\centering
\caption{The concentration of RIs in present NaI(Tl) crystals by the groups using NaI(Tl)
\cite{Bernabei2008, Amar_EPJ2019, Park2020}.
}
\label{tb:contami}
\begin{tabular}{l|rrrr} \hline
Conc. [$\mu$Bq/kg] & DAMA/LIBRA & COSINE & ANAIS & PICOLON \\ \hline
$^{40}$K & $<600$ & $515\sim1900$ & $540\sim 1200$ & $<480$ \\
$^{232}$Th & $2\sim 31$ & $2.5\sim35$ & $0.4\sim 4$ & $4.6\pm1.2$ \\
$^{226}$Ra & $8.7\sim124$ & $11\sim451$ & $2.7\sim 10$ & $8.7\pm1.5$ \\
$^{210}$Pb & $5\sim 30$ & $50\sim3800$ & $740\sim 3150$ & $28\pm5$ \\ \hline
\end{tabular}
\end{table}

Low background measurements are performed at the Kamioka Underground Observatory 
of Tohoku University in Hida City, Gifu Prefecture, Japan.
We installed the latest Ingot \# 94 in July 2021 inside a 10 cm thick lead and 5 cm 
thick oxygen-free copper shield. 
We flushed pure nitrogen to purge $^{222}$Rn in the shield.
NaI(Tl) immediately after the installation has much background due to
cosmogenic isotopes $^{125}$I ($T_{1/2}=59.408$~d) and $^{126}$I ($T_{1/2}=13.11$~d).
Since background due to these isotopes can be reduced by cooling for several months after installation, 
we started low-background measurement in January 2021.

\begin{figure}[h]
\centering
\includegraphics[width=\linewidth]{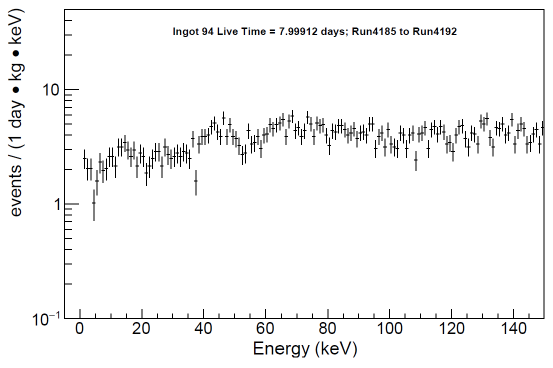}
\caption{Energy spectrum in low energy region taken by the newest ingot \# 94.}
\label{fg:lowe}
\end{figure}
The energy spectrum in the low energy region is shown in Figure \ref{fg:lowe}.
The background level between 5~\kevee \ and 10~\kevee \ was as low as 1.4 \dru.
Unfortunately, the noise level below 5~\kevee was unstable.
The noise events were mainly due to \v{C}erenkov radiation.
The second noise origin was a fluctuation in the baseline of the signal due to insufficient ground connection.
We are trying to remove these noise events by machine-learning.

\section{CANDLES Project}
\begin{figure}[h]
\centering
\includegraphics[width=\linewidth]{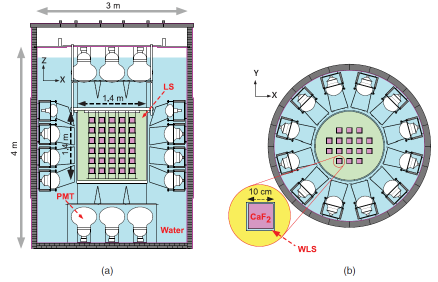}
\caption{The side view (a) and top view (b) of the CANDLES-III detector.
See details of the detector setup \cite{CANDLES2021}.}
\label{fg:candet}
\end{figure}
\begin{figure}[h]
\centering
\includegraphics[width=\linewidth]{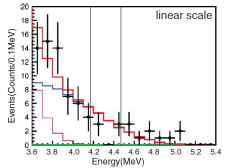}
\caption{The energy spectrum taken for the live time of 130.4~days \cite{CANDLES2021}.}
\label{fg:canene}
\end{figure}
CANDLES (CAlcium fluoride for studies of Neutrino and Dark matters by Low Energy Spectrometer) aims
to search for \znu \ of $^{48}$Ca by pure CaF$_{2}$ crystal \cite{CANDLES2021}.
We used 96 modules of pure CaF$_{2}$ crystals, where pure CaF$_{2}$ means an undoped CaF$_{2}$ crystal.
The dimension of each CaF$_{2}$ crystal is 10~cm cubic, and
the total mass of the $^{48}$Ca is 350~g.
The schematic drawing of the CANDLES-III detector is shown in Figure \ref{fg:candet}.

The experiment to search for \znu \ event was performed for the live time of 130.4~days.
We selected 21 pure CaF$_{2}$ modules from 96 ones.
The upper limit on the half-life of \znu \ was $5.6\times 10^{22}$~yr at 90~\% C.L.\@
The energy spectrum around the region of interest is shown in Figure \ref{fg:canene}.
We selected the best 21 crystals from 96 ones because of the higher contamination of radioactive isotopes.

The background energy spectrum was well reconstructed by known origins.
The background events consist of external gamma-rays from 
(n,$\gamma$) reaction in the surrounding rock
\cite{NAKAJIMA201854} and internal contamination (green line).
The severe internal contamination is observed from $^{208}$Tl and $^{212}$Bi-$^{212}$Po events\cite{CANDLES2021} 
around $Q$-value (blue line).
The \tnu \ events also an origin of background because of the poor enegy resolution (magenta line).
The red line gives the total amount of the known background events. 
We must reduce these contaminants lower than 10~$\mu$Bq/kg for 
further investigation of \znu.

\section{Discussion}
\subsection{Prospect of purification}
We discuss the prospect of further purification of inorganic crystals.
We have successfully purified NaI(Tl) crystal because the NaI is water-soluble and thus easy to process.
Water-insoluble inorganic crystals, such as CaF$_{2}$, have been applied in many double-beta decay experiments.
However, we must apply a different purification method to these crystals.
Previously, we tried to wash CaF$_{2}$ powder with nitric acid and remove adhered impurities, but this method cannot clean impurities inside the CaF$_{2}$ powder.

We started the purification work of CaF$_{2}$ as a water-insoluble inorganic crystal. 
Our target is less than 10~$\mu$Bq/kg for the uranium and thorium series.
In addition, we plan to enrich the target isotope $^{48}$Ca in CaF$_{2}$, 
so the purification method wastes as little calcium as possible during the purification.
After considering these conditions, we chose the method of synthesizing 
CaF$_{2}$ after purifying the raw materials of CaF$_{2}$.

The raw materials of CaF$_{2}$ are CaCl$_{2}$ and HF.\@
The HF has achieved an ultra-high purity of 12~N because it is used in the wet etching process of silicon wafers.
Therefore, we purified CaCl$_{2}$ by checking its purity.
 
 Before purification, we measured impurities in CaCl$_{2}$ by ICP-MS.\@
We bought CaF$_{2}$ powder from three providers, A, B, and C.\@
The concentration of $^{238}$U and $^{232}$Th were measured by Agilent 8900 at Tsukuba University.
The concentration of $^{238}$U was 166~ppt, 18~ppt, and 30~ppt in samples A, B, and C, respectively.
Unfortunately, the concentration of $^{232}$Th was not measured successfully because of the trouble in ICP-MS.\@

We tested the purification effectiveness by using a resin that catches uranium ions.
After repeating the purification process three times, the concentration of $^{238}$U was almost equal to that of the blank sample, 
confirming that the purification process was successful.
In the future, we will produce a high-purity, large (5~cm cubic) CaF$_{2}$ crystal and measure its impurities.

\subsection{Dark matter search by huge inorganic crystal}
Current NaI(Tl) detectors consist of small modules whose mass is around 10~kg. 
When an array of small modules is used, the background caused by components other than crystals, such as housings and light guides, increases and cannot be removed by event screening.
XENON-nT is a bulk detector and can remove external background events by measuring the location of events in the detector.

It is possible to select events in the center of the detector and remove the external background by making large inorganic crystals of 50~cm or larger.
Utilizing multiple PMTs gives the precise position of an event occurs.
The ratio of the detected number of scintillation photons enables the removal of the background events from surrounding materials.

\begin{figure}[ht]
\centering
\includegraphics[width=\linewidth]{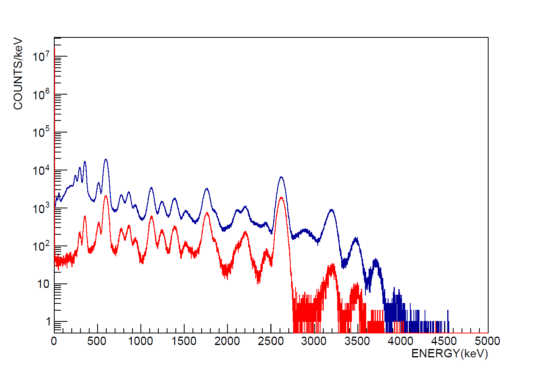}
\caption{Blue: Before external veto. Red: After external veto.}
\label{fg:diff}
\end{figure}
We estimated the reduction efficiency by installing an outer detector for an external veto by Monte Carlo simulation (Geant4).
The difference in the energy spectrum before and after the veto is shown in Figure \ref{fg:diff}.
We constructed a complex NaI(Tl) detector in this simulation.
The inner crystal was columnar shaped with 50~cm diameter and 50~cm length. 
The inner crystal was covered with 5~cm thick NaI(Tl) crystal.
A PMT was placed on the surface of the outer NaI(Tl) crystal and radioactive isotopes, $^{214}$Pb, $^{214}$Bi, and $^{208}$Tl were 
contained at the divider circuit of the PMT.\@

We removed the events in that the gamma rays interacted with inner and outer crystals.
The blue histogram is the original energy spectrum, and the red  
is the remaining events.
A reduction of almost two orders of magnitude was derived.
We are optimizing the detector design and estimating the expected sensitivity.

\section{Advantages and disadvantages of inorganic crystals}

\section{PICOLON project}

\section{Discussion}

\section{Acknowledgment}
We acknowledge the support of the Kamioka Mining and Smelting Company. This work was
supported by JSPS KAKENHI Grant No. 26104008, 19H00688, 20H05246, and Discretionary expense of the president of Tokushima University.
This work was also supported by the World Premier International Research Center Initiative (WPI Initiative).

\bibliographystyle{ptephy}
\bibliography{sample}

\begin{thebibliography}{10}

\bibitem{Sofue2020}
Y.~Sofue, Galaxies, {\bf 8}, 37 (2020).

\bibitem{Honma2012}
M.~Honma et~al., Publ.~Astron.~Soc.~Japan, {\bf 64}, 136 (2012).

\bibitem{Honma2015}
M.~Honma, T.~Nagayama, and N.~Sakai, Publ.~Astron.~Soc.~Japan, {\bf 67}, 70
  (2015).

\bibitem{Freese1988}
K.~Freese, J.~Frieman, and A.~Gould, Phys.~Rev.~\textbf{D}, {\bf 37},
  3388--3405 (1988).

\bibitem{PhysRevD.33.3495}
A~K. Drukier, K~Freese, and D~N. Spergel, Phys. Rev. D, {\bf 33}, 3495--3508
  (1986).

\bibitem{Peccei1977}
R.~D. Peccei and H.~R. Quinn, Phys.~Rev.~Lett., {\bf 38}, 1440--1443 (1977).

\bibitem{Weinberg1978}
Steven Weinberg, Phys.~Rev.~Lett., {\bf 40}, 223--226 (1978).

\bibitem{PhysRevLett.112.241302}
Agnese et~al., Phys. Rev. Lett., {\bf 112}, 241302 (2014).

\bibitem{CLINE2000373}
D~Cline et~al., Astroparticle Physics, {\bf 12}, 373--377 (2000).

\bibitem{PhysRevLett.129.161805}
E.~Aprile and Others, Phys. Rev. Lett., {\bf 129}, 161805 (2022).

\bibitem{EjiriJPSJ}
H~Ejiri, J.~Phys.~Soc.~Japan, {\bf 74}(8), 2101--2127 (2005).

\bibitem{Jones2021}
B~Jones,
\newblock The physics of neutrinoless double beta decay: A beginners guide,
\newblock In {\em Proceedings of Theoretical Advanced Study Institute 2020 "The
  Obscure Universe: Neutrinos and Other Dark Matters" - {TASI}2020
  {\textemdash} {PoS}({TASI}2020)}. Sissa Medialab (2021).

\bibitem{KL-Zen_PRL2023}
S.~Abe et~al., Phys. Rev. Lett., {\bf 130}, 051801 (2023).

\bibitem{CDMS2002}
D.~Abrams et~al., Phys.~Rev.~D, {\bf 66}, 122003 (2002).

\bibitem{CDMS2016}
R.~Agnese et~al., Phys. Rev. Lett., {\bf 116}, 071301 (2016).

\bibitem{AprileEpj2018}
E.~Aprile et~al., Eur.~Phys.~J.\textbf{C}, {\bf 78} (2018).

\bibitem{XENONnT2022}
E.~Aprile et~al., Phys. Rev. Lett., {\bf 129}, 161805 (2022).

\bibitem{Aprile2022}
E.~Aprile et~al., The European Physical Journal C, {\bf 82} (2022).

\bibitem{Park2020}
B.~J. Park et~al., Eur.~Phys.~J.~\textbf{C}, {\bf 80}, 814 (2020).

\bibitem{FushimiPTEP2020}
K.~Fushimi et~al., Prog.~Theor.~Exp.~Phys., ptab020 (2021).

\bibitem{FushimiISRD2016}
K.~Fushimi et~al., JPS~Conf.~Proc., {\bf 11}, 020003 (2016).

\bibitem{Bernabei2008}
R.~Bernabei et~al., Nucl.~Instru. and Meth. in Phys.~Res.~\textbf{A}, {\bf
  592}, 297--315 (2008).

\bibitem{Amar_EPJ2019}
J.~Amar{\'{e}} et~al., The European Physical Journal C, {\bf 79} (2019).

\bibitem{CANDLES2021}
S.~Ajimura et~al., Phys.~Rev.~D, {\bf 103}, 092008 (2021).

\bibitem{NAKAJIMA201854}
K.~Nakajima et~al., Astroparticle Physics, {\bf 100}, 54--60 (2018).

\end{thebibliography}

\end{document}